\documentclass[prc,twocolumn,epsfig,nofootinbib,floatfix,superscriptaddress]{revtex4-1}
\usepackage{graphics}
\usepackage{epsfig}
\usepackage{amsfonts}
\usepackage{amsmath}
\usepackage{float}
\usepackage{bm}
\usepackage{mathrsfs}
\usepackage{makecell}
\usepackage{siunitx}
\usepackage{color}
\usepackage{scrextend}
\usepackage{tablefootnote}
\usepackage[breaklinks=true,colorlinks=true,linkcolor=blue,urlcolor=blue,citecolor=red]{hyperref}

\begin{document}
	\title{Fine structure of the isoscalar giant monopole resonance in $^{58}$Ni, $^{90}$Zr, $^{120}$Sn and $^{208}$Pb}
	\author{A.~Bahini}
	\email[]{a.bahini@ilabs.nrf.ac.za}
	\affiliation{School of Physics, University of the Witwatersrand, Johannesburg 2050, South Africa}
	\affiliation{iThemba Laboratory for Accelerator Based Sciences, Somerset West 7129, South Africa}
	\author{P.~von~Neumann-Cosel}
	\email[]{vnc@ikp.tu-darmstadt.de}	
	\affiliation{Institut f\"{u}r Kernphysik, Technische Universit\"{a}t Darmstadt, D-64289 Darmstadt, Germany}
	\author{J.~Carter}
	\affiliation{School of Physics, University of the Witwatersrand, Johannesburg 2050, South Africa}	
	\author{I.~T.~Usman}
	\affiliation{School of Physics, University of the Witwatersrand, Johannesburg 2050, South Africa}	
	\author{N.~N.~Arsenyev}
	\affiliation{Bogoliubov Laboratory of Theoretical Physics, Joint Institute for Nuclear Research, 141980 Dubna, Russia}
	\author{A.~P.~Severyukhin}
	\affiliation{Bogoliubov Laboratory of Theoretical Physics, Joint Institute for Nuclear Research, 141980 Dubna, Russia}
	\author{E.~Litvinova}
	\affiliation{Department of Physics, Western Michigan University, Kalamazoo MI 49008-5252, USA}
    \affiliation{Facility for Rare Isotope Beams, Michigan State University, East Lansing, MI, 48824, USA}
    \affiliation{GANIL, CEA/DRF-CNRS/IN2P3, F-14076 Caen, France}
	\author{R.~W.~Fearick}
	\affiliation{Department of Physics, University of Cape Town, Rondebosch 7700, South Africa}			
	\author{R.~Neveling}
	\affiliation{iThemba Laboratory for Accelerator Based Sciences, Somerset West 7129, South Africa}	
	\author{P.~Adsley}
    \altaffiliation[Present address:]{ Department of Physics and Astronomy, Texas A\&M University, College Station, 77843-4242, Texas, USA and Cyclotron Institute, Texas A\&M University, College Station, 77843-3636, Texas, USA.}
	\affiliation{School of Physics, University of the Witwatersrand, Johannesburg 2050, South Africa}
	\affiliation{iThemba Laboratory for Accelerator Based Sciences, Somerset West 7129, South Africa}
	\affiliation{\hbox{Department of Physics, Stellenbosch University, Matieland 7602, Stellenbosch, South Africa}}
	\affiliation{\hbox{Irene Joliot Curie Lab, UMR8608, IN2P3-CNRS, Universit\'{e} Paris Sud 11, 91406 Orsay, France}}
	\author{N.~Botha}
	\affiliation{School of Physics, University of the Witwatersrand, Johannesburg 2050, South Africa}
	\author{J.~W.~Br\"{u}mmer}
	\affiliation{iThemba Laboratory for Accelerator Based Sciences, Somerset West 7129, South Africa}
	\affiliation{\hbox{Department of Physics, Stellenbosch University, Matieland 7602, Stellenbosch, South Africa}}
	\author{L. M. Donaldson}
	\affiliation{iThemba Laboratory for Accelerator Based Sciences, Somerset West 7129, South Africa}
	\author{S.~Jongile}
	\affiliation{iThemba Laboratory for Accelerator Based Sciences, Somerset West 7129, South Africa}
	\affiliation{\hbox{Department of Physics, Stellenbosch University, Matieland 7602, Stellenbosch, South Africa}}
	\author{T.~C.~Khumalo}
	\affiliation{School of Physics, University of the Witwatersrand, Johannesburg 2050, South Africa}
	\affiliation{iThemba Laboratory for Accelerator Based Sciences, Somerset West 7129, South Africa}	
	\affiliation{\hbox{Department of Physics, University of Zululand, Richards Bay 3900, South Africa}}			
	\author{M.~B.~Latif}	
	\affiliation{School of Physics, University of the Witwatersrand, Johannesburg 2050, South Africa}
	\affiliation{iThemba Laboratory for Accelerator Based Sciences, Somerset West 7129, South Africa}
	\author{K.~C.~W.~Li}
	\affiliation{iThemba Laboratory for Accelerator Based Sciences, Somerset West 7129, South Africa}	
	\affiliation{\hbox{Department of Physics, Stellenbosch University, Matieland 7602, Stellenbosch, South Africa}}
	\author{P.~Z.~Mabika}
	\affiliation{Department of Physics and Astronomy, University of the Western Cape, Bellville 7535, South Africa}
	\author{P.~T.~Molema}	
	\affiliation{School of Physics, University of the Witwatersrand, Johannesburg 2050, South Africa}
	\affiliation{iThemba Laboratory for Accelerator Based Sciences, Somerset West 7129, South Africa}	
	\author{C.~S.~Moodley}
	\affiliation{School of Physics, University of the Witwatersrand, Johannesburg 2050, South Africa}
	\affiliation{iThemba Laboratory for Accelerator Based Sciences, Somerset West 7129, South Africa}
	\author{S.~D.~Olorunfunmi}
	\affiliation{School of Physics, University of the Witwatersrand, Johannesburg 2050, South Africa}
	\affiliation{iThemba Laboratory for Accelerator Based Sciences, Somerset West 7129, South Africa}
	\author{P.~Papka}
	\affiliation{iThemba Laboratory for Accelerator Based Sciences, Somerset West 7129, South Africa}
	\affiliation{\hbox{Department of Physics, Stellenbosch University, Matieland 7602, Stellenbosch, South Africa}}
	\author{L.~Pellegri}
	\affiliation{School of Physics, University of the Witwatersrand, Johannesburg 2050, South Africa}
	\affiliation{iThemba Laboratory for Accelerator Based Sciences, Somerset West 7129, South Africa}
	\author{B.~Rebeiro}
	\affiliation{Department of Physics and Astronomy, University of the Western Cape, Bellville 7535, South Africa}
	\author{E.~Sideras-Haddad}
	\affiliation{School of Physics, University of the Witwatersrand, Johannesburg 2050, South Africa}
	\author{F.~D.~Smit}
	\affiliation{iThemba Laboratory for Accelerator Based Sciences, Somerset West 7129, South Africa}
	\author{S.~Triambak}
	\affiliation{Department of Physics and Astronomy, University of the Western Cape, Bellville 7535, South Africa}
	\author{M.~Wiedeking}
	\affiliation{School of Physics, University of the Witwatersrand, Johannesburg 2050, South Africa}
	\affiliation{iThemba Laboratory for Accelerator Based Sciences, Somerset West 7129, South Africa}
	\author{J.~J.~van~Zyl}
	\affiliation{\hbox{Department of Physics, Stellenbosch University, Matieland 7602, Stellenbosch, South Africa}}
	\date{\today}
	\begin{abstract}
		\noindent \textbf{Background:}
	    Over the past two decades high energy-resolution inelastic proton scattering studies were used to gain an understanding of the origin of fine structure observed in the isoscalar giant quadrupole resonance (ISGQR) and the isovector giant dipole resonance (IVGDR).~Recently, the isoscalar giant monopole resonance (ISGMR) in $^{58}$Ni, $^{90}$Zr, $^{120}$Sn and $^{208}$Pb was studied at the iThemba Laboratory for Accelerator Based Sciences (iThemba LABS) by means of inelastic $\alpha$-particle scattering at very forward scattering angles (including $\ang{0}$).~The good energy resolution of the measurement revealed significant fine structure of the ISGMR.
     
		\noindent\textbf{Objective:}~To extract scales by means of wavelet analysis characterizing the observed fine structure of the ISGMR in order to investigate the role of different mechanisms contributing to its decay width.
		 
		\noindent\textbf{Methods:}~Characteristic energy scales are extracted from the fine structure using continuous wavelet transforms.~The experimental energy scales are compared to different theoretical approaches
  performed in the framework of quasiparticle random phase approximation (QRPA) and beyond-QRPA including complex configurations using both non-relativistic and relativistic density functional theory.
		
		\noindent\textbf{Results:}~All models highlight the role of Landau fragmentation for the damping of the ISGMR especially in the medium-mass region.~Models which include the coupling between one particle-one hole (1p-1h) and two particle-two hole (2p-2h) configurations modify the strength distributions and wavelet scales indicating the importance of the spreading width.~The effect becomes more pronounced with increasing mass number.
		
		\noindent\textbf{Conclusions:}~Wavelet scales remain a sensitive measure of the interplay between Landau fragmentation and the spreading width in the description of the fine structure of giant resonances.~The case of the ISGMR is intermediate between the IVGDR, where Landau damping dominates, and the ISGQR, where fine structure originates from coupling to low-lying surface vibrations. 
	\end{abstract}

	\maketitle
\section{Introduction}
\label{s1}
Giant Resonances (GRs) as a collective mode of excitation are defined as small amplitude vibrations at high frequency (high $E_{\text{x}}$) around the ground state of the nucleus, involving most of the nucleons \cite{Harakeh}.~The isoscalar giant monopole resonance (ISGMR) was discovered four decades after the isovector giant dipole resonance (IVGDR) was first identified in the 1930s, and was later studied extensively at the Texas A\&M University (TAMU) Cyclotron Institute and the Research Center for Nuclear Physics (RCNP), through small angle (including $\ang{0}$) inelastic $\alpha$-scattering measurements at $240$ MeV and $386$ MeV, respectively.~However, only the gross properties (centroids and strengths in terms of exhaustion of sum rules) are so far reasonably well characterized and described by microscopic models \cite{GC_review2018}.~A systematic understanding of the widths, decay properties, and fine structure of the ISGMR remain largely unexplored topics. 
	
One of the main properties that define giant resonances is the width $\Gamma_{\text{GR}}$.~The width is as a result of the damping processes in the resonance, and has typical values of several MeV.~The damping of resonances can be described by different components as follows \cite{goeke1982theory}\\
	\begin{equation}
	\label{e1}
	\Gamma_{\text{GR}} = \Delta\Gamma + \Gamma^\uparrow + \Gamma^\downarrow \textrm{,}
	\end{equation}\\
with $\Delta\Gamma$ representing Landau damping which describes the fragmentation of the elementary one-particle one-hole ($1$p-$1$h) excitation, $\Gamma^\uparrow$ representing the escape width which corresponds to direct particle emission out of the continuum, and $\Gamma^\downarrow$ is the spreading width due to coupling to  two-particle two-hole ($2$p-$2$h) and many-particle many-hole (np-nh) states.~Information on the dominant damping mechanisms of nuclear giant resonances can be found in the properties and characteristics of the fine structure of the giant resonance.~This fine structure is the consequence of the mixture of multiple scales of fluctuations which are induced by the decay of nuclear states \cite{von2019electric}.~The spreading width $\Gamma ^\downarrow$ originates from the pre-equilibrium and statistical decay observed in compound nuclei.~Its stochastic coupling mechanism is well described by the doorway model \cite{lacroix1999}.
	
Through systematic studies at both the iThemba Laboratory for Accelerator Based Sciences (iThemba LABS) and RCNP, it was established that the main mechanism responsible for fine structure differs for different resonances.~In the case of the ISGQR it is due to coupling to low-lying surface vibrations 
\cite{shevchenko2004fine,shevchenko2009global,usman2011fine,usman2016fine,kureba2018wavelet}, but mainly due to Landau damping in the case of the IVGDR \cite{poltoratska2014,fearick2018origin,von2019electric,donaldson2020fine,carter2022}.~It is then of interest to know the mechanism leading to the fine structure in the case of ISGMR.~The present work aims at the investigation of the fine structure of ISGMR in $^{58}$Ni, $^{90}$Zr, $^{120}$Sn and $^{208}$Pb based on continuous wavelet analysis of high energy-resolution data extracted from $(\alpha,\alpha^\prime)$ reaction at very forward scattering angles.~The range of nuclei under investigation include singly- and doubly-magic nuclei and as such we opt to use theoretical approaches including degrees-of-freedom at and beyond the mean-field approximation of the quasiparticle random-phase approximation (QRPA).~In particular, we test calculations at the QRPA level (relativistic and non-relativistic) and beyond QRPA, allowing for the inclusion of correlated 2p-2h states by means of phonon-phonon coupling (PPC) employing Skyrme interactions and the relativistic quasiparticle time blocking approximation (RQTBA) developed for relativistic energy density functionals.
	
   \section{Experiment and data analysis}
	\label{s2}
\begin{figure} 
		\centering
		\includegraphics[trim=0.7cm -0.5cm 0 2.8cm, scale=0.48]{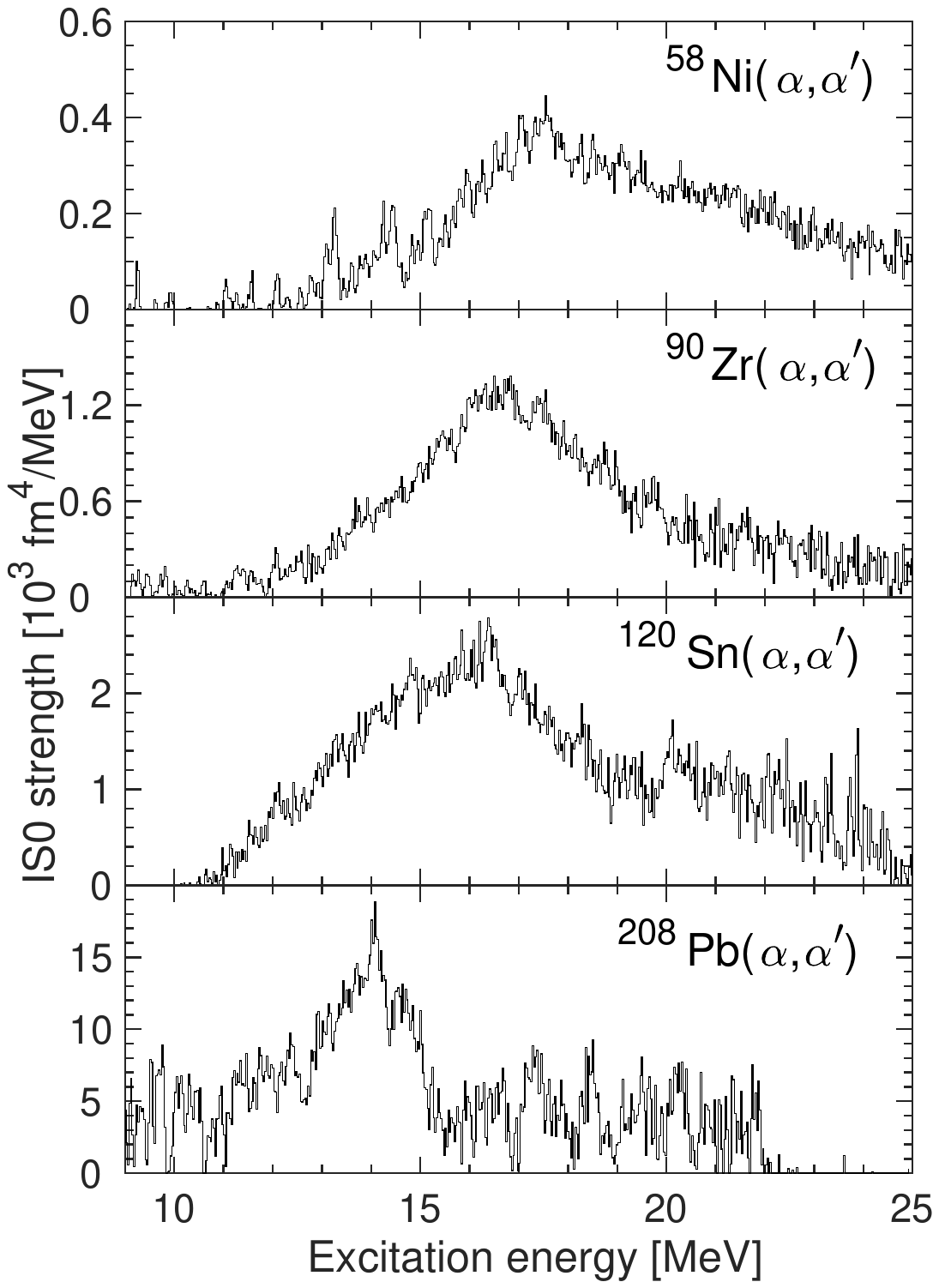}
		\caption{Isoscalar monopole strength distributions obtained with the ($\alpha,\alpha^\prime$) reaction at $E_\alpha = 196$ MeV on $^{208}$Pb,$^{120}$Sn,$^{90}$Zr and $^{58}$Ni. See text for details.}
		\label{FIG:1}
	\end{figure}

The details of the experimental procedure followed in this study are given in Ref.~\cite{Armand2_PRC2022}.~As such, only the main points are summarized here.~The experiment was performed at the Separated Sector Cyclotron (SSC) facility of iThemba LABS, South Africa.~A beam of $196$ MeV $\alpha$-particles was inelastically scattered off self-supporting $^{58}$Ni, $^{90}$Zr, $^{120}$Sn and $^{208}$Pb targets with areal densities ranging from $0.7$ to $1.4$ mg/cm$^2$ and isotopically enriched to values $>96\%$.~The reaction products were momentum analyzed by the K$600$ magnetic spectrometer positioned at laboratory scattering angles $0^{\circ}$ and $4^{\circ}$ \cite{nev11}.~Following extraction of the inelastic scattering cross sections, the isoscalar monopole (IS0) strength distributions were obtained by means of the Difference-of-Spectra (DoS) technique with excitation energy-dependent corrections (see Ref.~\cite{Armand2_PRC2022} for details).~The correction factors used here are based on the multipole decomposition analysis of $L > 0$ cross sections in previous experiments at RCNP \cite{nayak2006,gupta2018isoscalar,li2010isoscalar,patelphd}.~The resulting spectra shown in Fig.~\ref{FIG:1}, binned to $30$ keV,  demonstrate significant fine structure up to excitation energies of approximately $20$ MeV.

\begin{figure} 
		\centering
		\includegraphics[scale=0.62]{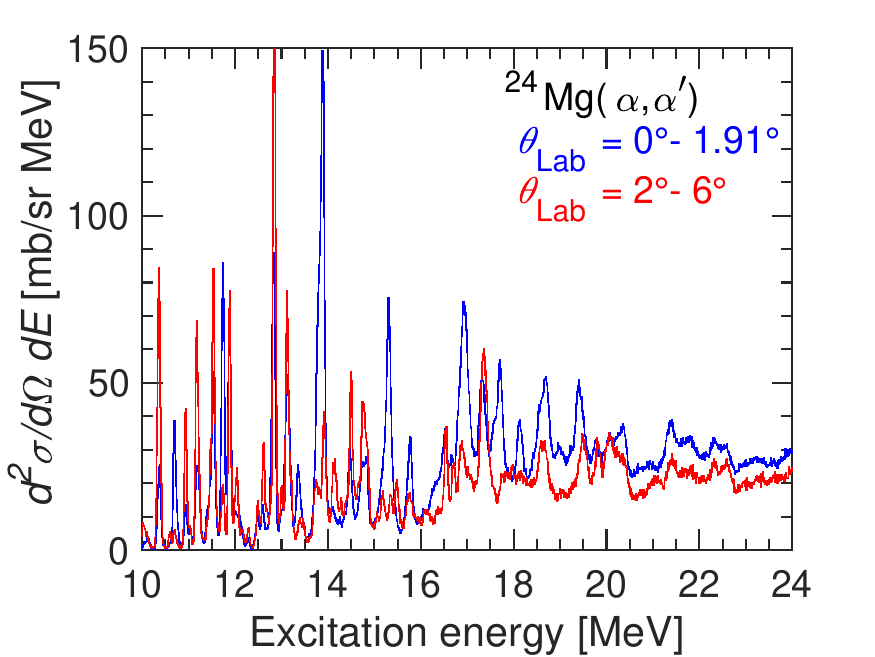}
		 \vspace{0.2cm}\caption{Double-differential cross sections measured for the $^{24}$Mg($\alpha$,$\alpha^\prime$) reaction at $E_\alpha = 196$ MeV for the angular range $\theta_{\text{Lab}} = 0^{\circ} - 1.91^{\circ}$ (blue) and $\theta_{\text{Lab}} = 2^{\circ} - 6^{\circ}$ (red).}
		\label{FIG:2a}
	\end{figure}

The momentum calibration for both the zero- and four-degrees measurements was very important in order to ensure that no false structures are induced in the difference spectrum of the DoS metod.~This was achieved using well-known states in $^{24}$Mg \cite{kaw2013,bor1981} as shown in Fig.~\ref{FIG:2a}.~An energy resolution of $\approx 70$ keV full width at half maximum  (FWHM) was obtained for both the zero- and four-degree measurements.

	\section{Theoretical models}
	\label{s3}
In the following we discuss the four models that will be used to provide IS0 strength functions to be compared with the experimental results.	

	\subsection{Non relativistic approaches with a Skyrme interaction}
	\label{s31}

One of the successful tools for nuclear structure studies is the quasiparticle random phase approximation (QRPA) with the self-consistent mean-field derived by making use of the Skyrme interaction.~Such QRPA calculations do not require new parameters since the residual interaction is derived from the same energy density functional~(EDF) as that determining the mean-field.~The residual interaction in the particle-hole channel and in the particle-particle channel can be obtained as the second derivatives of the EDF with respect to the particle density and the pair density, respectively.~To build the QRPA equations on the basis of Hartree-Fock (HF) Bardeen-Cooper-Schrieffer (BCS) quasiparticle states with the residual interaction is a standard procedure \cite{Terasaki2005}.~The wave functions of the ground state is the QRPA phonon vacuum $|0\rangle$ and the one-phonon QRPA states given by $Q_{\lambda\mu{i}}^{+}|0\rangle$ have energy $\omega_{\lambda{i}}$, where the index $\lambda$ denotes the total angular momentum and the index $\mu$ is its $z$-projection in the laboratory system.~The dimensions of the QRPA matrix grow rapidly with the size of the nucleus.~Using the finite-rank separable approximation~(FRSA) \cite{Giai1998} for the residual interactions, the eigenvalues of the QRPA equations can be obtained as the roots of a relatively simple secular equation \cite{Severyukhin2008}.~It enables us to perform QRPA calculations in very large two-quasiparticle spaces.~The cut-off of the discretized continuous part of the single-particle (SP) spectra is at the energy of $100$~MeV.~This is sufficient to exhaust practically all the energy-weighted sum rule.~Because of this large configurational space, we do not need effective charges.~We use the Skyrme-EDF SLy4 \cite{Chabanat1998} with a nuclear matter incompressibility modulus 
$K_{\infty}$=229.9 MeV.~It is worth to mention that the SLy4 set provides a good description of the ISGMR in medium- and heavy-mass spherical nuclei \cite{Khan2009,Severyukhin2017a,Arsenyev2021}.~The pairing correlations were generated by a surface peaked density-dependent zero-range force, and the pairing strength was taken as $-870$ MeV{}fm$^{3}$ \cite{Severyukhin2017b,Arsenyev2021}.~To limit the pairing SP space, we used a smooth cutoff at $10$~MeV above the Fermi energies \cite{Severyukhin2008}.~In the QRPA solution, there exists the problem of the spurious $0^{+}$~state which can appear at low energy ($<2$~MeV).~It is shown that the spurious state is very well separated from the physical modes~\cite{Li2008} and we can thus ignore them.

The qualitative agreement with high energy-resolution experimental data can only be achieved by including phonon-phonon coupling (PPC) effects, such as the fragmentation of the QRPA states \cite{donaldson2020fine}.~We follow the basic ideas of the quasiparticle-phonon model (QPM) \cite{Soloviev1992}.~Using the completeness and orthogonality conditions for the phonon operators one can express bifermion operators through the phonon ones and the Hamiltonian can be rewritten in terms of quasiparticle and phonon operators, see Ref.~\cite{Severyukhin2004}.~This method has already been introduced in Refs.~\cite{Severyukhin2004,Severyukhin2012}.~We construct the wave functions from a linear combination of one- and two-phonon configurations as\\
\begin{equation}
  \begin{split}
  \Psi_{\nu}(\lambda\mu)=\Biggl(&\sum\limits_{i}R_{i}(\lambda\nu)Q_{\lambda\mu i}^{+}\\
  +\sum\limits_{\lambda_{1}i_{1}\lambda_{2}i_{2}}
  P_{\lambda_{2}i_{2}}^{\lambda_{1}i_{1}}(\lambda\nu)&
  \left[Q_{\lambda_{1}\mu_{1}i_{1}}^{+}Q_{\lambda_{2}\mu_{2}i_{2}}^{+}\right]_{\lambda\mu}
  \Biggr)|0\rangle~,
  \label{wf2ph}
  \end{split}
\end{equation}\\
where the $[\ldots]_{\lambda\mu}$ stands for angular momentum coupling. Using the variational principle one obtains a set of linear equations for the amplitudes $R_{i}(\lambda\nu)$ and $P_{\lambda_{2}i_{2}}^{\lambda_{1}i_{1}}(\lambda\nu)$\\
\begin{equation}
  \begin{split}
  &(\omega_{\lambda{i}}-E_{\nu})R_{i}(\lambda\nu)\\
  &+\sum\limits_{\lambda_{1}i_{1}\lambda_{2}i_{2}}
  U_{\lambda_{2}i_{2}}^{\lambda_{1}i_{1}}(\lambda{i})
  P_{\lambda_{2}i_{2}}^{\lambda_{1}i_{1}}(\lambda\nu)=0~,
  \label{2pheq1}
  \end{split}
\end{equation}

\begin{equation}
  \begin{split}
  &\sum\limits_{i}U_{\lambda_{2}i_{2}}^{\lambda_{1}i_{1}}(\lambda{i})
  R_{i}(\lambda\nu)\\
  &+2(\omega_{\lambda_{1}i_{1}}+\omega_{\lambda_{2}i_{2}}-E_{\nu})
  P_{\lambda_{2}i_{2}}^{\lambda_{1}i_{1}}(\lambda\nu)=0~.
  \label{2pheq2}
  \end{split}
\end{equation}\\
For its solution it is required to compute the Hamiltonian matrix elements coupling one- and two-phonon configurations \cite{Severyukhin2004,Severyukhin2012}\\
\begin{equation}
  U_{\lambda_{2}i_{2}}^{\lambda_{1}i_{1}}(\lambda{i})=
  \langle0|Q_{\lambda{i}}H\left[Q_{\lambda_{1}i_{1}}^{+}
  Q_{\lambda_{2}i_{2}}^{+}\right]_{\lambda}|0\rangle~.
\label{U2ph}
\end{equation}\\
The rank of the set of linear equations (\ref{2pheq1}) and (\ref{2pheq2}) is equal to the number of one- and two-phonon configurations included in the wave functions Eq.~(\ref{wf2ph}).~Equations (\ref{2pheq1}) and (\ref{2pheq2}) have the same form as the QPM equations \cite{Soloviev1992,Voronov2000}, but the SP spectrum and the parameters of the residual interaction are calculated with the Skyrme EDF.~Our calculation is based on the QRPA formulation.~It should be noted as well that the ground state correlations beyond the QRPA~\cite{Voronov2000,Severyukhin2015} may play an important role. In this context the problem of convergence and stability of solutions of the beyond QRPA models and the so-called problem of double counting have been discussed in \cite{Tselyaev2013}.~However, all these questions are beyond the scope of the present paper, and require separate studies.

In the present study, to construct the wave functions of the excited $0^{+}$ states we take all the two-phonon configurations below $25$~MeV into account that are built from the QRPA phonons with multipolarities $\lambda^{\pi}=0^{+}$, $1^{-}$, $2^{+}$, $3^{-}$, $4^{+}$ and $5^{-}$ coupled to $0^{+}$.~It is interesting to examine the energies and reduced transition probabilities of the lowest $2^{+}$, $3^{-}$, and $4^{+}$ RPA states, which are the important ingredients for understanding the nature of the two-phonon $0^{+}$~states of $^{208}$Pb.~The results of the RPA calculation for the energies, the $B(E\lambda)$ values, and the structure of these states are given in Table~\ref{tabRPA}. Note that the energies and the reduced transition probabilities calculated within the FRSA are very close to those calculated in the RPA with a full treatment of the Skyrme-type p{-}h residual interaction~\cite{Colo2013}.~As one can see, 
the overall agreement of the energies and $B(E\lambda)$ values with the experimental data~\cite{Heisenberg1982,Spear1983} looks reasonable.~The overestimates regarding energies indicate that there is a room for the PPC effects (see for example \cite{Severyukhin2004}). 

\begin{table}
\caption{Energies, transition probabilities, and structures of the
RPA low-lying states in $^{208}$Pb. The two-quasiparticle configuration
contributions greater than 5{\%} are given. Experimental
data are taken from Refs.~\cite{Heisenberg1982,Spear1983}.} \label{tabRPA}
\begin{ruledtabular}
\begin{tabular}{ccccccc}
 $\lambda^{\pi}_{1}$&\multicolumn{2}{c}{Energy}&\multicolumn{2}{c}{$B(E\lambda;0^{+}_{gs}\rightarrow\lambda^{\pi}_{1})$}&Structure\\
          &\multicolumn{2}{c}{(MeV)} &\multicolumn{2}{c}{(e$^2$b$^{\lambda}$)} &\\
          & Expt.  &  Theory & Expt.        &  Theory         &\\
\noalign{\smallskip}\hline\noalign{\smallskip}
 $2^{+}_{1}$&4.085 & 5.2    & $0.318{\pm}0.016$& 0.34 &54{\%}$\{2g\frac{9}{2},{1i\frac{13}{2}}\}_{\nu}$\\
            &      &         &                  &      &36{\%}$\{2f\frac{7}{2},{1h\frac{11}{2}}\}_{\pi}$\\
            &      &         &                  &      &5{\%}$\{1h\frac{9}{2},{1h\frac{11}{2}}\}_{\pi}$\\
\noalign{\smallskip}
 $3^{-}_{1}$&2.615 & 3.6    & $0.611{\pm}0.012$& 0.93 &13{\%}$\{2g\frac{9}{2},{1p\frac{3}{2}}\}_{\nu}$\\
            &      &         &                  &      &9{\%}$\{1i\frac{11}{2},{2f\frac{5}{2}}\}_{\nu}$\\
            &      &         &                  &      &7{\%}$\{1j\frac{15}{2},{1i\frac{13}{2}}\}_{\nu}$\\
            &      &         &                  &      &21{\%}$\{1h\frac{9}{2},{2d\frac{3}{2}}\}_{\pi}$\\
            &      &         &                  &      &9{\%}$\{1i\frac{13}{2},{1h\frac{11}{2}}\}_{\pi}$\\
            &      &         &                  &      &9{\%}$\{2f\frac{7}{2},{3s\frac{1}{2}}\}_{\pi}$\\
\noalign{\smallskip}
 $4^{+}_{1}$&4.323 & 5.6    & $0.155{\pm}0.011$& 0.15 &33{\%}$\{2g\frac{9}{2},{1i\frac{13}{2}}\}_{\nu}$\\
            &      &         &                  &      &41{\%}$\{1h\frac{9}{2},{1h\frac{11}{2}}\}_{\pi}$\\
            &      &         &                  &      &15{\%}$\{2f\frac{7}{2},{1h\frac{11}{2}}\}_{\pi}$\\
\end{tabular}
\end{ruledtabular}
\end{table}

The rank of the set of linear equations~(\ref{2pheq1},\ref{2pheq2}) is equal to the number of the one- and two-phonon configurations included in the wave functions.~This means that the two-phonon configurational space is now enlarged by the phonon compositions
$[\lambda_{1}^{\pi_{1}}{\otimes}\lambda_{2}^{\pi_{2}}]_{\text{QRPA}}$, i.e.,
$[0^{+}{\otimes}0^{+}]_{\text{QRPA}}$, $[1^{-}{\otimes}1^{-}]_{\text{QRPA}}$,
$[2^{+}{\otimes}2^{+}]_{\text{QRPA}}$, $[3^{-}{\otimes}3^{-}]_{\text{QRPA}}$,
$[4^{+}{\otimes}4^{+}]_{\text{QRPA}}$ and $[5^{-}{\otimes}5^{-}]_{\text{QRPA}}$.~As an example, for $^{208}$Pb, in the case of the set SLy4, the PPC calculation takes into account $40$ monopole phonons, $49$ dipole phonons, $74$ quadrupole phonons, $109$ octupole phonons, $93$ hexadecapole phonons and $104$ pentapole phonons when all the one- and two-phonon configurations below $25$~MeV are included.~It is worth mentioning that the major contribution to the ISGMR strength distribution is brought about by the coupling between the $[0^{+}]_{\text{RPA}}$ and $[3^{-}{\otimes}3^{-}]_{\text{RPA}}$ components \cite{Severyukhin2016}. 

The IS0 strength function is computed as\\
\begin{equation}
\text{IS0}(\omega)=\sum\limits_{\nu}\left|\langle0^{+}_{\nu}|\hat{M}_{\lambda=0}|0^{+}_{\text{g.s.}}\rangle\right|^2
 \rho(\omega-E_{\nu})~,
\end{equation}\\
where $\left|\langle0^{+}_{\nu}|\hat{M}_{\lambda=0}|0^{+}_{\text{g.s.}}\rangle\right|^2$
is the transition probability of the $\nu$-th $0^{+}$ state.~The transition operator of the ISGMR is defined as\\
\begin{equation}
\hat{M}_{\lambda=0}=
\sum\limits_{i=1}^{A}r^2_i~.
\end{equation}\\
The IS0 strength function is averaged out by a Lorentzian distribution with a smoothing parameter of $\Delta$ as follows\\
\begin{equation}
 \rho(\omega-E_{\nu})=\frac{1}{2\pi}\frac{\Delta}{(\omega-E_{\nu})^2+\Delta^{2}/4}~.
\end{equation}\\
For accurate comparison between theoretical and experimental results, a smoothing parameter equivalent to the experimental energy resolution is used.~The strength is then summed over the appropriate number of bins.~The inclusion of the PPC lead to small down shifts of the centroid energy of the ISGMR.~It is worth mentioning that the first systematical Skyrme-EDF study of the influence of the quasiparticle-vibration coupling on the ISGMR centroid has been done in~\cite{Li2023}.

\subsection{Relativistic approaches with an effective meson-exchange interaction}
\label{s33}

Two relativistic self-consistent approaches, the relativistic quasiparticle random phase approximation (RQRPA) and the relativistic quasiparticle time blocking approximation (RQTBA), were employed to compute the isoscalar monopole response in the nuclear systems under study.~RQRPA pioneered in Ref. \cite{Paar2003} is confined by two-quasiparticle ($2q$) configurations interacting via the exchange of mesons between nucleons.~The effective interaction is a derivative of the self-consistent mean field with respect to the nucleonic density, i.e., both are defined by the same set of eight parameters NL3$^{\ast}$, namely the nucleon-meson coupling constants and meson masses.~The latter values are slightly refitted compared to their vacuum values, and non-linear couplings of the scalar meson are adopted to obtain a realistic mean field, whereas the compressibility modulus $K_{\infty} = 258$ MeV corresponds to this parameter set \cite{Lalazissis2009}.~In most cases, RQRPA reasonably describes the collective states at both low and high energies, however, it is known to overestimate the centroid of the giant monopole resonance in nuclei lighter than lead.

Many details of the nuclear spectra are stipulated by much more complex wave functions of the excited states than the $2q$ ones.~The leading approximation beyond (R)QRPA includes $2q\otimes phonon$ configurations, which produce considerable fragmentation of the (R)QRPA modes and generate much richer spectral structures.~In the relativistic framework, this approach was first formulated and implemented numerically as the relativistic quasiparticle time blocking approximation in Ref.~\cite{LRT2008}, where it was derived from the phenomenological nucleon-phonon self-energy by the time blocking technique following Ref.~\cite{Tselyaev1989}.~Later, the time blocking was ruled out as an unnecessary step when the response theory is derived from an \textit{ab-initio} Hamiltonian in the model-independent equation of motion (EOM) framework  \cite{LitvinovaSchuck2019,LitvinovaZhang2022}.~In the EOM formalism, RQTBA was obtained as one of the possible approaches to the dynamical kernel, or in-medium long-range interaction, which keeps the leading effects of emergent collectivity.~The developments of Refs.~\cite{Litvinova2015,LitvinovaSchuck2019,LitvinovaZhang2022,Litvinova2023a} also allowed for a self-consistent  extension of the theory to the $2q\otimes 2phonon$ configurations, i.e., to the three-particle-three-hole level, which produces further refinement of the spectral strength distributions.

In Refs.~\cite{LitvinovaSchuck2019,LitvinovaZhang2022} it was shown that all the many-body models for the fermionic response are derivable from the exact \textit{ab-initio} theory.~The QRPA, or one-phonon, approach follows when completely neglecting the EOM's dynamical kernel for the response function and considering only the $2q$ configurations.~The dynamical kernel, which generates complex configurations beyond the $2q$ ones, couples to the hierarchy of EOMs of growing complexity and may be approximated by a cluster decomposition to make the many-body problem tractable.~The minimal truncation on the two-body level leads to the quasiparticle-vibration coupling and multiphonon approaches, depending on the desirable correlation content, which is expressed by Eq. (60) of Ref.~\cite{LitvinovaZhang2022}.~Using an effective interaction instead of the bare interaction between the nucleons requires the subtraction \cite{Tselyaev2013}, which eliminates the double counting of the complex configurations from the effective interaction, thereby recovering the consistency of the theory.

 Both the original and extended versions of RQTBA have demonstrated a good performance in the description of nuclear excited states in both neutral \cite{LRT2008,EndresLitvinovaSavranEtAl2010,LanzaVitturiLitvinovaEtAl2014} and charge-exchange \cite{RobinLitvinova2016,RobinLitvinova2018,Robin2019,Scott2017} channels, showing remarkable improvements with respect to RQRPA.~Most notably, the $2q\otimes phonon$ configurations produce a reasonable degree of fragmentation of the $2q$-states already in the leading approximation.~In particular, the description of the low-energy (soft) modes was refined considerably, which is especially important for the applications to \textit{r}-process nucleosynthesis in stellar environments and supernovae evolution \cite{LRW2020,LR2021}.~The so-called nuclear fluffiness puzzle was addressed recently in Ref.~\cite{Litvinova2023} within the same approach applied to the ISGMR in various nuclei across the nuclear chart.~It was shown that the self-consistent relativistic response theory, including $2q\otimes phonon$ configurations beyond RQRPA, can reasonably describe both the centroids and the widths of the ISGMR in the lead, tin, zirconium, and nickel isotopes.~Reference \cite{Litvinova2023} was the major stepping stone on the way to consensus between a softer equation of state extracted from the compressibility of finite nuclei and a stiffer one required by recent analyses of neutron star data.
  

In this work, we employ the same version of RQTBA as in Ref.~\cite{Litvinova2023} with pairing correlations taken into account on an equal footing with the quasiparticle-vibration coupling in terms of the $2q\otimes phonon$ configurations, which are included up to $50$ MeV.~The corresponding amplitudes are generated from the characteristics of the relativistic mean-field quasiparticles and RQRPA phonons in a parameter-free way.~The phonon model space is truncated using the same criteria as in the series of earlier calculations, for instance, in Refs.~\cite{EL2016,Litvinova2023}.~The complete set of the $2q$ configurations was included in the calculations, which allows for maximal suppression of the spurious component.~The subtraction procedure, following Ref.~\cite{Tselyaev2013}, eliminates the $2q\otimes phonon$ contributions from the effective interaction to avoid their double counting, ensures converged results within the given configuration space, and preserves the decoupling of the spurious state.~The imaginary part of the energy variable in the response function, corresponding to half of the width of the resulting peaks, is chosen to be $\Delta = 35$ keV to match the experimental energy resolution of $70$ keV. 

 Fragmentation of the ISGMR due to the $2q\otimes phonon$ configurations included in RQTBA was found in reasonable agreement with the lower-resolution data of Refs.~\cite{GC_review2018,Li2007} and \cite{Gupta2016} for  $^{208}$Pb, $^{120}$Sn and $^{80}$Zr, respectively.~An accurate comparison was performed and discussed in Ref.~\cite{Litvinova2023}, where also the ISGMR's centroid shift due to these configurations was investigated and linked to quadrupole collectivity, which is typically enhanced in soft mid-shell nuclei.~The fragmentation of the monopole response is overall weaker than that of the higher multipoles, and both the fragmentation and centroid position are sensitive to the details of the numerical scheme, such as the basis completeness and self-consistency.~The latter is stipulated using the same effective interaction in the static and dynamic sectors and the subtraction procedure.

\section{Fine-Structure analysis}
\label{s4}

Different methods can be employed in order to gain insight into the characteristic energy-scales of the fine structure of giant resonances, such as the entropy index method \cite{aiba2003fluctuation}, a multifractal analysis \cite{lacroix2000multiple}, or a method based on the Continuous Wavelet Transform (CWT) \cite{shevchenko2004fine}.~The CWT method was used previously in the analysis of the fine structure observed in the ISGQR \cite{shevchenko2009global,usman2011fine,kureba2018wavelet} and the IVGDR \cite{donaldson2020fine,carter2022}, and will therefore also be employed in this study.~A brief summary of the formalism and techniques of the wavelet analysis,  discussed in detail elsewhere \cite{donaldson2020fine}, is provided here.



\subsection{Wavelet-analysis formalism}
\label{s41}

Wavelet analysis is an effective tool to analyze multiscale structures \cite{resnikoff2012wavelet}.~Fourier analysis can also play the same role through superposition of sine and cosine functions to analyze periodic signals.~However, the sinusoidal functions used to represent data are non-local and infinite, this then makes the Fourier analysis inappropriate in the case of the fine structure analysis of giant resonances.~Wavelet analysis offers information on the localization of high-frequency signal aspects \cite{gra1995introduction}.~In addition, wavelet analysis is not constrained to the usage of sinusoidal functions only.~These features together allow a study of the evolution of the frequency pattern of a given signal with optimized resolution.~Another useful feature of the wavelet analysis is the approximation of any background contribution in the experimental data, through the so-called \textit{vanishing moments} of the wavelet function.

The choice of a wavelet plays an important role when performing wavelet analysis.
The most frequently used functions for wavelet analysis are discussed in Ref.~\cite{shevchenko2008analysis}.~The detector response of the magnetic spectrometer used in the experiments is well approximated by a Gaussian line shape.~As such, for the analysis of the fine structure, the Morlet wavelet consisting of a Gaussian envelope on top of a periodic structure is the most suitable.~The Morlet wavelet is given by \cite{mallat1998wavelet}\\
\begin{equation}
\label{e271a}
\Psi(x) = \dfrac{1}{\pi^{\frac{1}{2}}f_\text{b}}\exp\left(2\pi if_\text{c} - \frac{x^2}{f_\text{b}}\right)~,
\end{equation}\\	
where $f_\text{b} = 2$ and $f_\text{c} = 1$ are used as the wavelet bandwidth and the center frequency of the wavelet, respectively.~This wavelet-function or \textit{wavelets} must meet a set of requirements:
\begin{itemize}
\item the function oscillating with a mean value that equals zero and
\item the function must have finite length.
\end{itemize}
Mathematically, the above conditions can be written as\\
\begin{equation}
\int_{-\infty}^{\infty} \Psi^*(x)dx = 0
\end{equation}\\
and\\
\begin{equation}
K_{\Psi} = \int_{-\infty}^{\infty} \mid \Psi^2(x)\mid dx < \infty ~,
\end{equation}\\
where $\Psi(x)$ is a real or complex function used as mother-wavelet with $\Psi^*(x)$ as its complex conjugate.~Here, $K_{\Psi}$ is the norm of the wavelet.~The second condition defines the local feature of wavelets.~The Continuous Wavelet Transform and the Discrete Wavelet Transform (DWT) are the two categories of wavelets transforms available.~Their main properties and the comparison between the two transforms are discussed in Ref.~\cite{shevchenko2008analysis}.~For the purposes of the present analysis, only the application of the CWT will be discussed.

The convolution of a given signal $\sigma(E)$ with the wavelet function (generally complex-conjugated) yields the coefficients of the wavelet transform.~This is explicitly given by \cite{mallat1998wavelet}\\
\begin{equation}
\label{e2611}
C(\delta E, E_\text{x}) = \dfrac{1}{\sqrt{\delta E}}\int \sigma(E)\Psi^*\left(\dfrac{E_\text{x} - E}{\delta E}\right) dE~,
\end{equation}\\
where $C(\delta E, E_\text{x})$ are the coefficients of the wavelet transform, $\delta E$ represents the bin size and moreover a scaling factor.~The parameter $E_\text{x}$ shifts the position of the wavelet  across the excitation-energy range, hence allowing access to the scale-localization information.~The parameters $\delta E$ and $E_\text{x}$ are varied continuously in the framework of CWT.~The values of the coefficients indicate to what extent the form of the scaled and shifted wavelet is close to the original spectrum. 

The extraction of wavelet energy-scales can be achieved from the wavelet coefficient plot as peaks in the corresponding power spectrum.~The wavelet power spectrum is the projection of the summed squared wavelet coefficients onto the wavelet scale axis\\
\begin{equation}
\label{e2612}
P_\omega(\delta E) = \dfrac{1}{N}\sum_{i }\mid C_i(\delta E)C_i^*(\delta E)\mid~,
\end{equation}\\
where $P_\omega(\delta E)$ represents the power as a function of scale $\delta E$ summed at each scale value over the index $i = N$ with $N$ the number of energy bins in the excitation-energy region considered.

\subsection{Application of the CWT on the ISGMR data}

\begin{figure}
	\includegraphics[trim=1.3cm -0.5cm 0 0.6cm,scale=0.48]{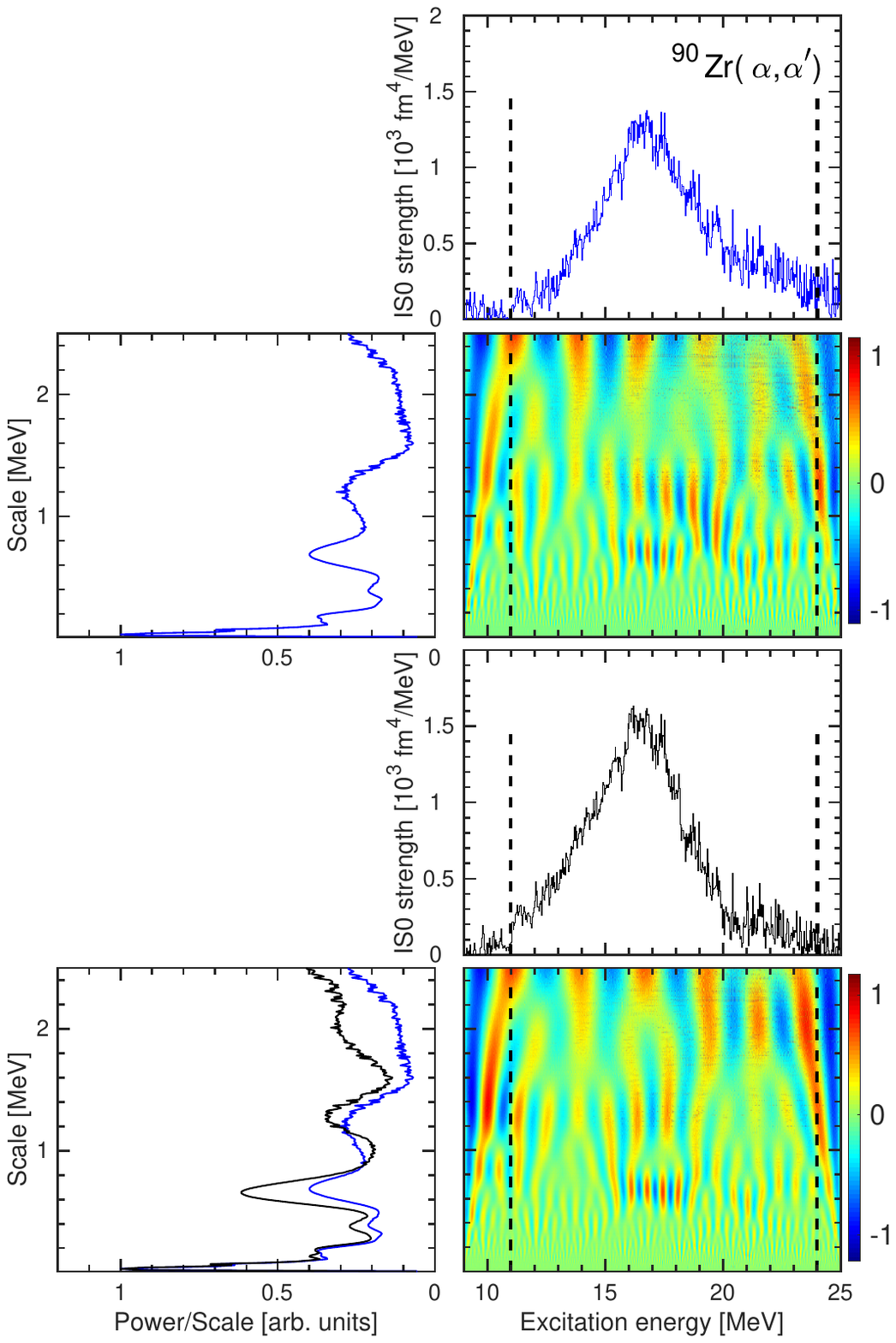}
 		\caption{Top set (right column): IS0 strength of $^{90}$Zr obtained using the RCNP-based energy-dependent correction factor as determined in Ref.~\cite{Armand2_PRC2022}.~Top set (lower right): Density plot of the real part of the CWT coefficients of the data.~Top set (left column):~Corresponding power spectrum for the excitation-energy region indicated by the vertical dashed lines ($11~ \text{MeV} \leq E_{\text{x}} \leq 24~ \text{MeV}$) in the top right plot.~Bottom set: Same as the top set but for the difference spectrum obtained using TAMU-based energy-dependent correction factors.~Bottom set (left column): The corresponding power spectrum shown in black, contrasted with the power spectrum from the top set (blue line).}
 		\label{FIG:3}
\end{figure}

The wavelet analysis was performed following the techniques outlined above.
A CWT was used to generate the wavelet coefficients Eq.~(\ref{e2611}) as a function of excitation energy, for each of the IS0 strength distributions of the nuclei under investigation.~In Ref.~\cite{Armand2_PRC2022}, it was discussed that the IS0 strength distributions extracted with the DoS technique need to be corrected by energy-dependent factors determined from the MDA analysis of previous experiments on the same nuclei.~It is, therefore, important to investigate the impact of this dependency on the fine structure analysis.

The sensitivity of the wavelet analysis to the different correction factors is illustrated in Fig.~\ref{FIG:3} for the case of $^{90}$Zr.~IS0 strength distributions obtained with correction factors derived from Refs.~\cite{gupta2018isoscalar} and \cite{krishichayan2015g} are shown in the top and third row, respectively.~The two-dimensional plots of the wavelet coefficients are displayed in the second and fourth panels on the right-hand side of Fig.~\ref{FIG:3}.~The intermittent appearance of blue (red) regions indicating negative (positive) values, result from the oscillatory structure of the mother wavelet (Eq.~\ref{e271a}) used in the analysis.~Extracted wavelet coefficients are then projected onto the scale axis to generate the power spectrum shown in the two panels on the left-hand side of Fig.~\ref{FIG:3}.~These spectra display the distribution of the scales in the excitation-energy region chosen for the analysis.~The presence of characteristic scales is indicated by peaks and points of inflection in the power spectra.

When comparing the power spectra resulting from the IS0 strength distributions with different correction factors, it is clear that even though there are relative power changes, very similar scale energies are found.~The details of the analysis techniques used in Ref.~\cite{Armand2_PRC2022} do, therefore, not affect the extraction of information on the fine structure of the GMR extracted with wavelet techniques.~All results presented in the next section are based on the DoS results that employed the correction factors based on RCNP experiments \cite{nayak2006,gupta2018isoscalar,li2010isoscalar,patelphd}.

\section{Damping of the ISGMR - wavelet energy-scales comparison}
\label{s5}
\begin{figure} 
		\centering
		\includegraphics[trim=0.8cm -2cm 0.2 2.5cm, scale=0.43
  ]{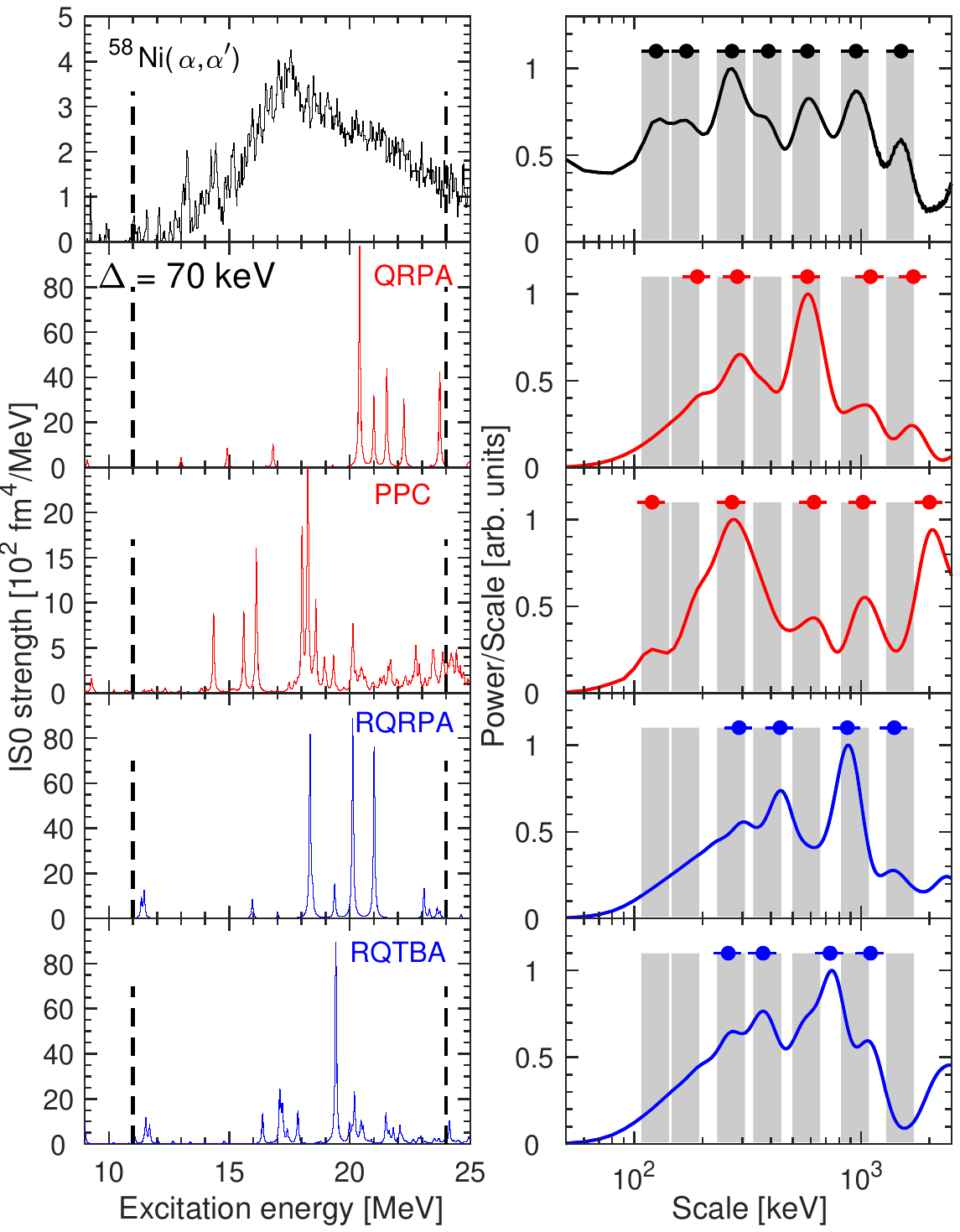}
		\caption{Left column: Experimental IS0 strength in $^{58}$Ni (top row) in comparison with model predictions (rows 2-5) folded with the experimental energy resolution.~The vertical dashed lines indicate the summation region of the wavelet coefficients ($11 - 24$ MeV) to determine the power spectra.~Right column: Corresponding power spectra. Scales are indicated by filled circles with the associated errors, and for the experimental results additionally by vertical grey bars.
  }
		\label{FIG:4}
	\end{figure}

\begin{figure}  
		\centering
		\includegraphics[ trim=0.8cm -2cm 0.2 2.5cm, scale=0.43]{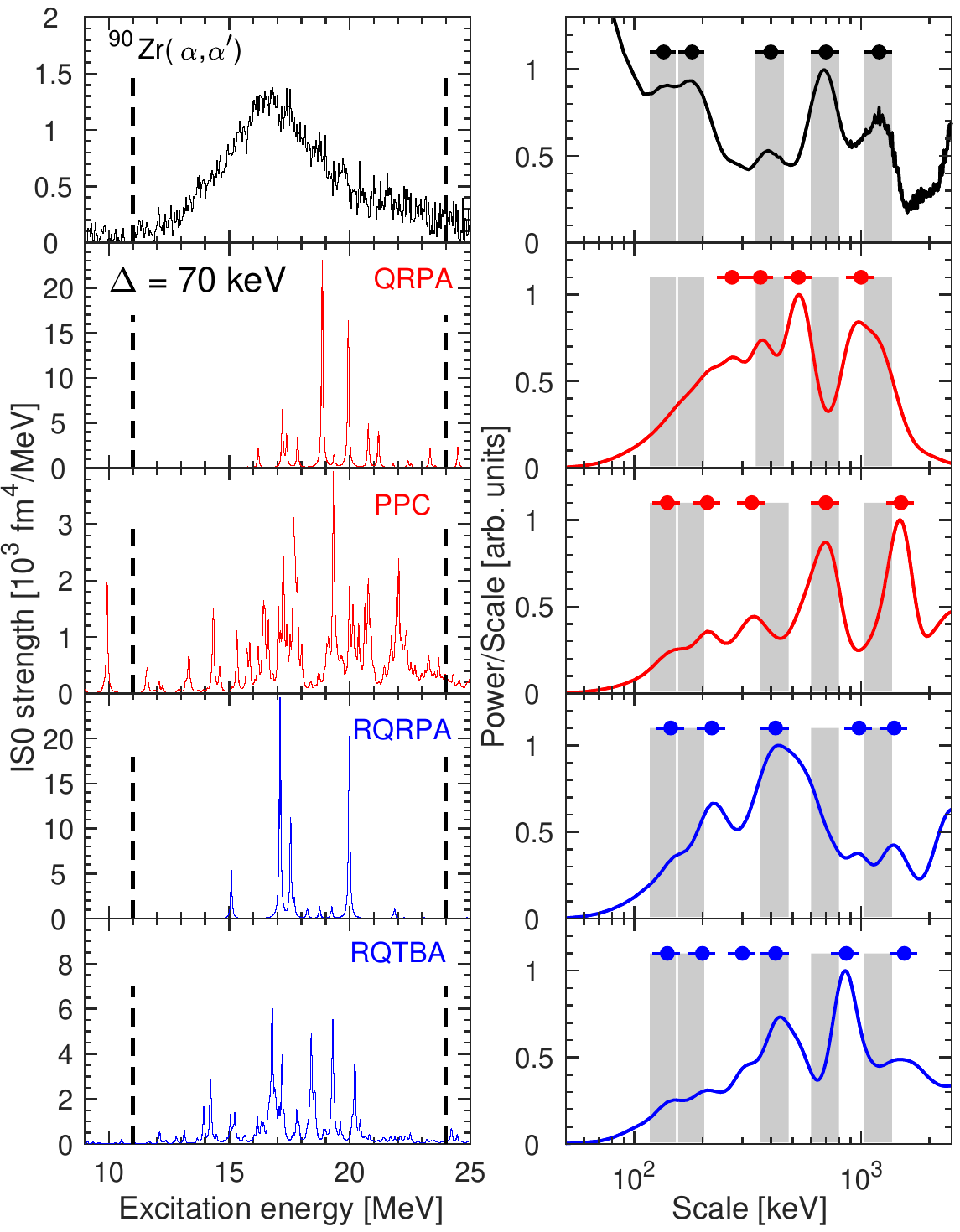}
		\caption{Same as Fig.~\ref{FIG:4}, but for $^{90}$Zr.}
		\label{FIG:5}
	\end{figure}

\begin{figure}  
		\centering
		\includegraphics[ trim=0.8cm -2cm 0.2 2.5cm, scale=0.43]{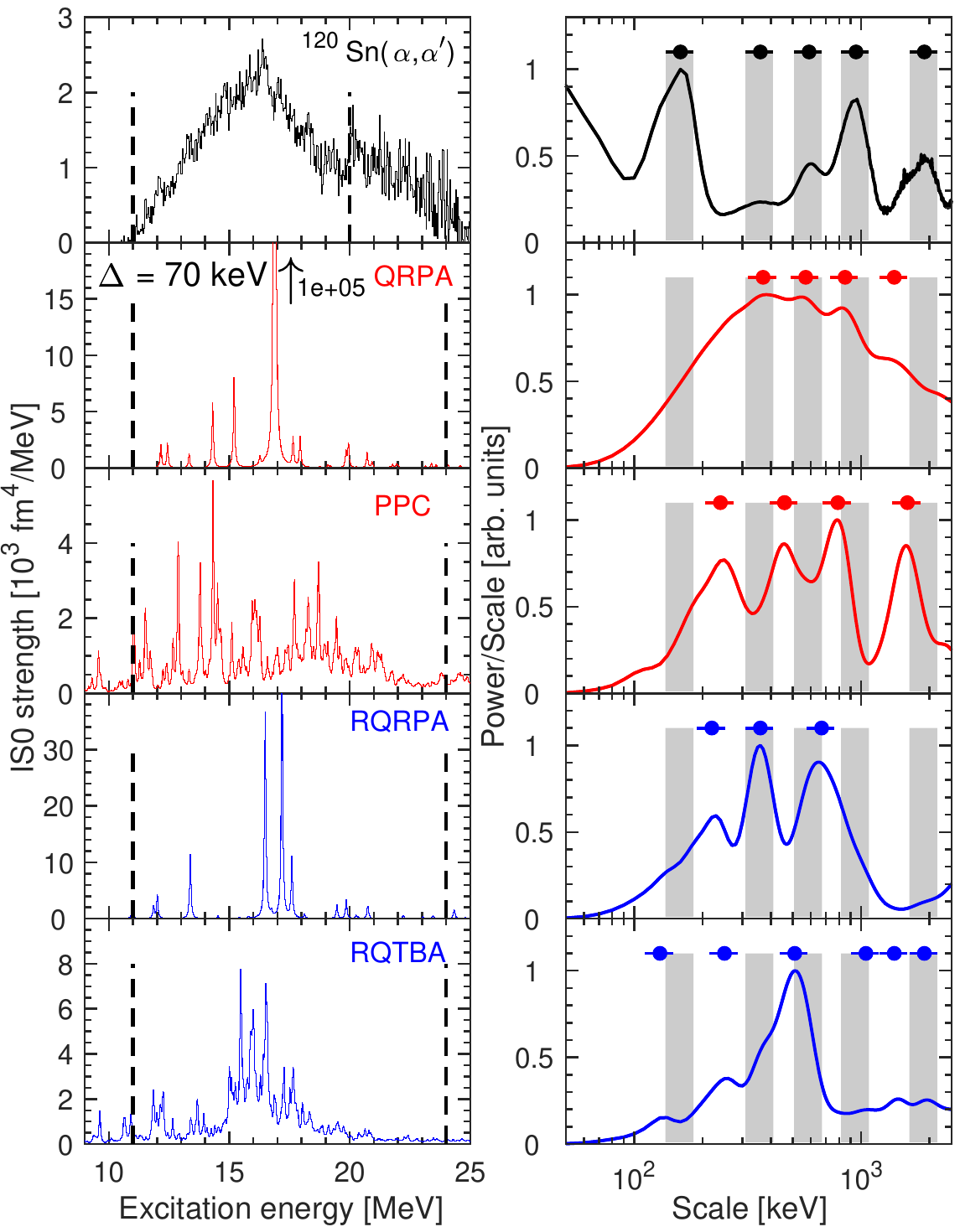}
		\caption{Same as Fig.~\ref{FIG:4}, but for $^{120}$Sn and the excitation-energy region from $11$ to $20$ MeV (experimental data).}
		\label{FIG:6}
	\end{figure}

\begin{figure}  
		\centering
		\includegraphics[ trim=0.8cm -2cm 0.2 2.5cm, scale=0.43]{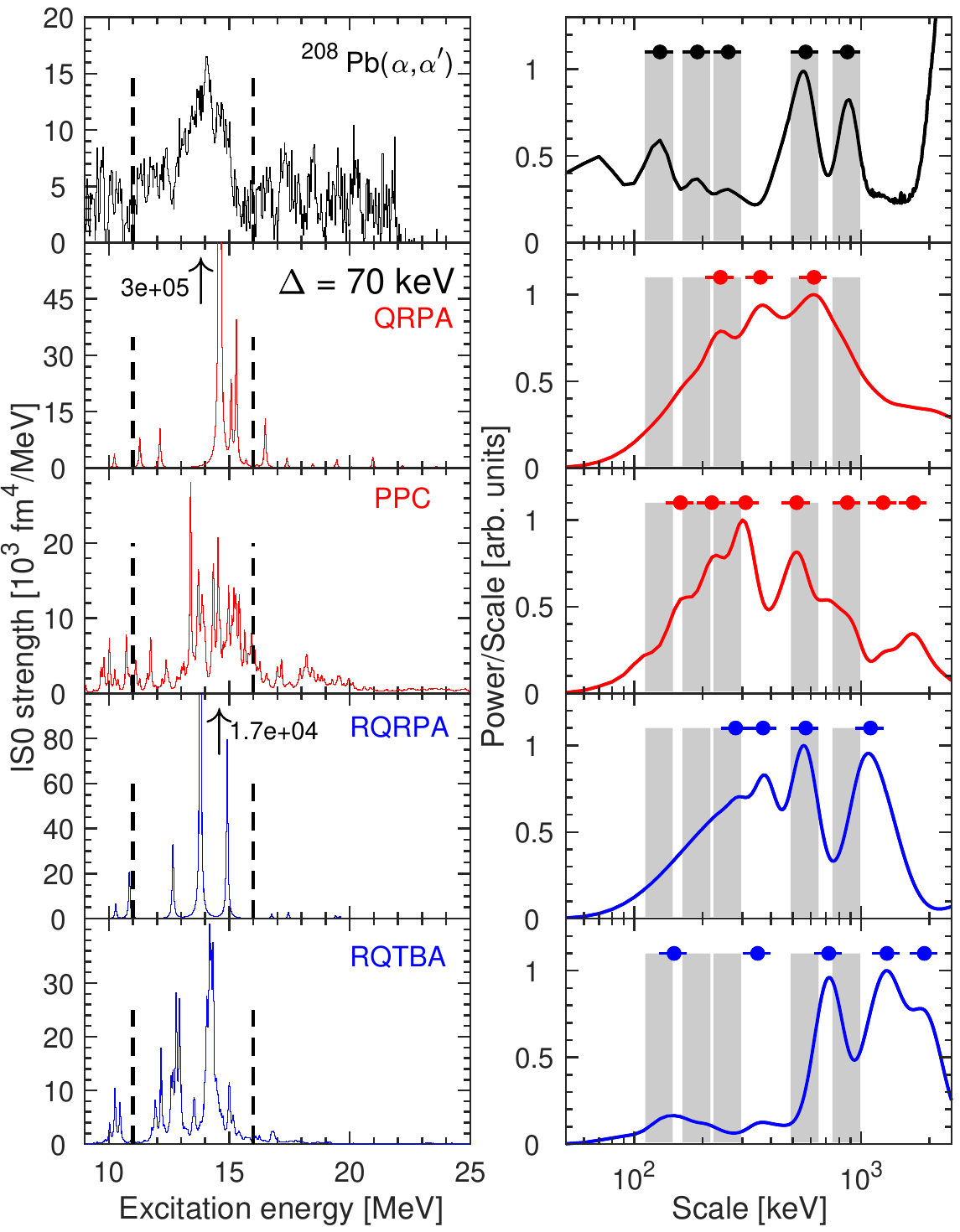}
		\caption{Same as Fig.~\ref{FIG:4}, but for $^{208}$Pb and the excitation-energy region from $11$ to $16$ MeV over which the wavelet coefficients were summed in order to determine the corresponding power spectra.}
		\label{FIG:7}
	\end{figure}

In this section, the results of the wavelet analysis of the experimental and model IS0 strength functions are presented.~They are summarized in Figs.~\ref{FIG:4} - \ref{FIG:7}.~For each nucleus, different energy regions have been considered for the analysis depending on the location of the main ISGMR peak.~These regions are indicated by the vertical dashed lines shown in panels on the left-side of Figs.~\ref{FIG:4} - \ref{FIG:7}.~Characteristic scales are extracted from the power spectra and displayed as black (experiment), red (QRPA and PPC) and blue (RQRPA and RQTBA) filled circles.~The associated error is given by one standard deviation of the corresponding width-like scale corresponding to half of the peak width (FWHM), cf.~\cite{carter2022}.~For comparison purposes and in order to facilitate the determination of similar scales in the corresponding power spectra from the model calculations, the results obtained from experiments are also displayed as vertical grey bars in all right-side panels of Figs.~\ref{FIG:4} - \ref{FIG:7}.~For the sake of better display, their widths have been reduced to $2/3$ of the standard deviation.~The extracted energy scales, both experimental and theoretical, are also listed in Tables \ref{table:5} - \ref{table:8}. ~When two scales agree within error, they are placed in the same column to ease comparison between experiment and model results.


\subsection{General observations}

Before entering a detailed discussion for each studied nucleus, we summarize some general observations when comparing experimental and theoretical strength distributions and wavelet scales.~Both theoretical approaches describe the energy centroids of the ISGMR in a similar way with a slight overestimation (about $1$ MeV) for the lighter nuclei $^{58}$Ni and $^{90}$Zr and a good reproduction for the heavier cases $^{120}$Sn and $^{208}$Pb.~We note that a shift between experimental and theoretical centroids does not impact on the CWT.~The inclusion of complex configuration leads to an increased fragmentation, but effects are much stronger in the PPC than in the RQTBA calculations.~In fact, except for $^{58}$Ni, the PPC results resemble the experimental widths quite well.

Characteristic scales deduced from the fine structure are significantly modified when going from QRPA level to inclusion of two-phonon or $2q\otimes phonon$ configurations.~In most cases additional scales appear in overall better agreement with the number of scales extracted from the experimental data.~The capability to reproduce absolute scale energies varies from case to case as discussed below.~The smallest scale with values $130-160$ keV is prominent in the power spectra of all studied nuclei, but generally much weaker in the theoretical results.~Consistent with findings in the IVGDR, this scale is an exclusive signature of the spreading width since it only appears in calculations with inclusion of complex configurations.

\subsection{$^{58}$Ni}

\begin{table}
	\caption{
 Energy scales extracted for $^{58}$Ni in the excitation energy region $11~ \text{MeV} \leq E_{\text{x}} \leq 24~ \text{MeV}$. 
 Equivalent characteristic energy-scale values are vertically aligned.}	
	\label{table:5}
	\begin{center}
		\setlength{\arrayrulewidth}{0.5pt}
		\setlength{\tabcolsep}{0.13cm}
		\renewcommand{\arraystretch}{1.2}	
		\begin{tabular}{cccccccccc}
			\hline\hline
			\multicolumn{10}{l}{Dataset \hspace{3cm}Scales (keV)}\\
			\hline
			Expt. & $130$ & $170$ & $270$ & $390$ & $580$ & 
			& $950$ & $1500$ & 
			\\
			QRPA &  & $190$ & $290$ &  & $580$ & & $1100$
			& $1700$ & 
			\\
  		    PPC & $120$ &  & $270$ &  & $620$ & 
			& $1020$ &  & $2000$
			\\ 
            RQRPA &  &  & $290$ & $420$ &  & 
			& $870$ & $1400$ & 
			\\
            RQTBA &  &  & $260$ & $370$ &  & $730$
			& $1100$ &  & 
			\\
			\hline\hline	
		\end{tabular}
	\end{center}	
\end{table}

The CWT of the experimental IS0 strength distribution shows the largest number of scales $(7)$ of the four nuclei studied.~The numbers observed for QRPA and RQRPA are $5$ and $4$, respectively, and no additional scales appear when complex configurations are included.~The major experimental scales at $270$ and $950$ keV are reproduced by all models, while the scale at $580$ keV is only seen by the QRPA/PPC approach (eventually also shifted to $730$ keV in RQTBA).

A scale $>1$ MeV is seen in all but the RQTBA result.~Indeed, this scale is observed in the RQRPA result because of Landau fragmentation into a few main states, while the RQTBA result exhibits a single prominent peak only.~Finally, a small scale found to be a generic feature of coupling to $2q\otimes phonon$ configurations in previous studies of the IVGDR and ISGQR is visible in the PPC result only.

\subsection{$^{90}$Zr}

\begin{table}
	\caption{Energy scales extracted for $^{90}$Zr in the excitation energy region $11~ \text{MeV} \leq E_{\text{x}} \leq 24~ \text{MeV}$. Equivalent characteristic energy-scale values are vertically aligned.}	
	\label{table:6}
	\begin{center}
		\setlength{\arrayrulewidth}{0.5pt}
		\setlength{\tabcolsep}{0.15cm}
		\renewcommand{\arraystretch}{1.2}	
		\begin{tabular}{ccccccccc}
			\hline\hline
			\multicolumn{9}{l}{Dataset \hspace{3cm}Scales (keV)}\\
			\hline
			Expt. & $135$ & $180$ &  & $400$ &  & $700$ & $1200$ &
			\\
			QRPA &  &  & $270$ & $360$ & $530$ &  & $1000$ &
			\\
  		    PPC & $140$ & $210$ &  & $330$ &  & $700$ &  & $1500$
			\\ 
            RQRPA & $145$ & $220$ &  & $400$ &  &  & $980$ $1400$ &
			\\
            RQTBA & $140$ & $200$ & $300$ & $420$ &  & $850$ &  & $1500$
			\\
			\hline\hline	
		\end{tabular}
	\end{center}	
\end{table}

A significant effect of the coupling to complex configurations is seen for $^{90}$Zr in all models.~The number of scales is increased from $4$ to $5$ (PPC), respectively $5$ to $6$ (RQTBA).~The PPC and RQTBA results can account for all experimental scales below $1$ MeV including the observation of two scales at small energies ($\leq 200$ keV).
An additional weaker scale at $300$ keV not seen in the data is predicted by the RQTBA approach.~A larger scale $> 1$ MeV consistent with the experimental one at $1200$ keV is found in both models but the predicted value ($1500$ keV) is somewhat large. 

\subsection{$^{120}$Sn}
 
\begin{table}
	\caption{Energy scales extracted for $^{120}$Sn in the excitation energy region $11~ \text{MeV} \leq E_{\text{x}} \leq 20~(24)~\text{MeV}$ for experimental data (theoretical calculations). Equivalent characteristic energy-scale values are vertically aligned.}	
	\label{table:7}
	\begin{center}
		\setlength{\arrayrulewidth}{0.5pt}
		\setlength{\tabcolsep}{0.17cm}
		\renewcommand{\arraystretch}{1.2}	
		\begin{tabular}{ccccccccc}
			\hline\hline
			\multicolumn{9}{l}{Dataset \hspace{3cm}Scales (keV)}\\
			\hline
			Expt. & $160$ &  & $360$ &  & $590$ & $950$ &  & $1900$
            \\
			QRPA &  &  & $370$ &  & $570$ & $850$ & $1400$ & 
			\\
  		    PPC &  & $240$ &  & $460$ &  & $790$ &  & $1600$
			\\ 
            RQRPA &  & $220$ & $360$ &  & $670$ &  &  & 
			\\
            RQTBA & $130$ & $250$ &  &  & $510$ & $1050$ & $1400$ & $1900$ 
			\\
			\hline\hline	
		\end{tabular}
	\end{center}	
\end{table}

The experimental summation window for the wavelet power has been reduced to $11 -20$ MeV since the strength at higher excitation-energies might be attributed to a less than perfect subtraction of the low-energy ﬂank of the ISGDR that dominates the background cross sections \cite{Uchida2003}.~The $5$ experimental scales are to be compared with $4$ in the PPC approach (with no change from the QRPA result despite a considerable increase of fragmentation of the strength distribution) and $6$ in RQTBA ($3$ in RQRPA).~RQTBA also accounts well for the absolute scale values except one ($250$ keV vs.~$360$ keV experimentally) and an additional weak scale at $1400$ keV not seen in the data.
The PPC scales below $1$ MeV are systematically shifted to higher values as compared with experiment.

\subsection{$^{208}$Pb}

\begin{table}
	\caption{Energy scales extracted for $^{208}$Pb in the excitation energy region $11~ \text{MeV} \leq E_{\text{x}} \leq 16~ \text{MeV}$. Equivalent characteristic energy-scale values are vertically aligned.}	
	\label{table:8}
	\begin{center}
		\setlength{\arrayrulewidth}{0.5pt}
		\setlength{\tabcolsep}{0.15cm}
		\renewcommand{\arraystretch}{1.2}	
		\begin{tabular}{cccccccccc}
			\hline\hline
			\multicolumn{9}{l}{Dataset \hspace{3cm}Scales (keV)}\\
			\hline
			Expt. & $130$ & $190$ & $260$ &  & $570$ & $870$ &  & 
			\\
			QRPA  &  &  & $240$ & $360$ & $620$ &  &  & 
			\\
  		    PPC &  $160$ & $220$  & $310$ &  & $520$ & $870$ & $1250$ & $1700$
			\\ 
            RQRPA &  &  & $280$ & $370$ & $570$ &  & $1100$ & 
			\\
            RQTBA & $150$ &  &  & $350$ &  & $720$ & $1300$ & $1900$
			\\
			\hline\hline	
		\end{tabular}
	\end{center}	
\end{table}


Because of the problem of remaining ISGDR strength in the DoS subtraction \cite{Uchida2003} already mentioned for $^{120}$Sn, the wavelet power summation is restricted to $11 - 17$ MeV.~Although the same window is used for the theoretical results, this might affect the power spectrum, in particular at larger scale values.
Thus, the discussion here is restricted to scales $< 1$ MeV.

Reverse to the $^{120}$Sn case, inclusion of complex configurations increases the number of scales in the PPC approach to $5$ in accordance with experiment, while it remains at $3$ when going from RQRPA to RQTBA.~The PPC result quantitatively reproduces all scale values within the typical uncertainties.~RQTBA reproduces the smallest and largest scale (in the scale region up to $1$ MeV).

\section{Conclusions and Outlook}

In this study, we present high energy-resolution IS0 strength distributions over a wide mass range extracted from measurements of the ($\alpha,\alpha^\prime$) reaction at $196$ MeV and extreme forward-scattering angles (including $\ang{0}$), revealing significant fine structure.~Characteristic energy scales were extracted from a Continuous Wavelet Transform (CWT) analysis of the data to investigate the role of Landau fragmentation and spreading width in the damping of the ISGMR.

The experimental results are compared to microscopic calculations of the ISGMR strength functions based on the QRPA and beyond-QRPA using both non-relativistic and relativistic density functional theory.~The extracted experimental energy scales are well reproduced by the models where in most cases a number of scales can be approximately reproduced, but the one-to-one correspondence varies from case to case.

The wavelet scales remain a sensitive measure of the interplay between Landau fragmentation and spreading width in the description of the fine structure of giant resonances \cite{von2019electric}.~In the case of the ISGMR, Landau damping is prominent in the medium-mass region while the spreading width increases with mass number and makes the largest contribution in heavy nuclei.~The relative importance of both contributions is intermediate between the IVGDR, where Landau damping dominates over the spreading width even for heavy nuclei, and the ISGQR, where fine structure is entirely due to coupling to low-lying surface vibrations (except maybe for light nuclei). 

The fragmentation of the $J = 0$ response is generally weaker than that of the $J>0$ one because of the smaller amount of the $2q \otimes phonon$ or $phonon\otimes phonon$ configurations allowed by the angular momentum conservation.~Furthermore, both the fragmentation and centroid position are sensitive to the details of the numerical scheme, such as the basis completeness and self-consistency.~The latter is stipulated using the same effective interaction in the static and dynamic sectors and the subtraction procedure.~One question to be addressed in future work is the impact of the subtraction procedure on the PPC approach considering the different degree of fragmentation with respect to the RQTBA results.
 
A complete response theory for atomic nuclei should include continuum, unnatural parity and isospin-flip phonons, complex ground-state correlations, and higher-order configurations, which are expected to further affect the fine structure of the strength functions and improve the description of the characteristic energy scales.~These effects are beyond the scope of this work and will be addressed by future efforts.

\section*{ACKNOWLEDGEMENTS}
The authors thank the Accelerator Group at iThemba LABS for the high-quality dispersion-matched beam provided for this experiment.~This work was supported by the National Research Foundation (NRF) of South Africa (Grant No.~$85509$, $118846$ and $129603$), the Deutsche Forschungsgemeinschaft under contract SFB $1245$ (Project ID No.~$79384907$), as well as an NRF-JINR grant (JINR200401510986).~A.B. acknowledges financial support through iThemba LABS and the NRF of South Africa.~P.A. acknowledges support from the Claude Leon Foundation in the form of a postdoctoral fellowship.~E.L. acknowledges support by the GANIL Visitor Program and funding from the National Science Foundation of the United States of America US-NSF under the US-NSF CAREER Grant PHY-1654379 and US-NSF Grant PHY-2209376.~N.N.A. acknowledges support from the Russian Science Foundation (Grant No.~RSF-21-12-00061).
	
\end{document}